\documentstyle[aps]{revtex}
\begin{document}

\title{Dynamical identification of open quantum systems}
\author{H. Mabuchi}
\address{Norman Bridge Laboratory of Physics 12-33, California Institute
of Technology, Pasadena, CA 91125, USA}
\date{August 12, 1996}
\maketitle

\begin{abstract}
I propose a quantum trajectories approach to parametric identification
of the effective Hamiltonian for a Markovian open quantum system, and
discuss an application motivated by recent experiments in cavity quantum
electrodynamics.  This example illustrates a strategy for quantum
parameter estimation that efficiently utilizes the information
carried by correlations between measurements distributed in time.
\end{abstract}

\bigskip
\noindent
The rapidly-developing theory of quantum parameter estimation (QPE) focuses
on the design of optimal measurement strategies for extracting information
about c-number parameters ${\bf \vec{\theta}}$ that characterize a given
quantum system.  While much progress has been made in applying QPE to the
parametric identification of {\em prepared quantum states}, little or no
attention has yet been paid to the problem of estimating parameters that
characterize a {\em dynamical quantum system}.  In this letter I consider
the latter aspect of QPE within the context of quantum optics, and describe
a quantum trajectories method for estimating parameters that appear in the
effective Hamiltonian for a Markovian open quantum system.

Recent theoretical work in QPE~\cite{Brau92a,Sand95,Wise95} has focused on
a paradigm in which an experimenter is provided with one or more copies
of a quantum state $\rho_0$ drawn from a single-parameter family ${\bf\rho}
(\theta)$, and is asked to determine the value $\theta_0$ such that
$\rho_0={\bf\rho}(\theta_0)$.  The experimenter knows the form of ${\bf\rho}
(\theta)$ and can make arbitrary measurements on the states she is given, but
does not know the value of $\theta_0$ {\it a priori}.  In such situations one
can actually derive a mathematical representation of the optimal quantum
measurement for the purpose of estimating $\theta_0$, and optimize over all
possible statistical reductions of the measurement results as
well~\cite{Brau92b}.  Accordingly, there exists a generalized version of the
Cram\'{e}r-Rao inequality~\cite{Hole82} that establishes a fundamental bound
on the rate of convergence for estimators based on repeated measurements whose
marginal statistics are fully determined by a single, unchanging density
matrix~\cite{Leon95}.  My purpose here is to consider a related but distinct
aspect of QPE, namely the estimation of parameters appearing in the equations
of motion that govern the time-evolution of a quantum system.  In this paradigm,
which is closely related to that of classical {\em system
identification}~\cite{Ljun87}, the hypothetical experimenter wishes to determine
which system model ${\cal H}_0$ in a parametrized family ${\bf\cal H}({\vec
\theta})$ best accounts for the dynamical behavior of a given quantum system.
In contrast to the conventional QPE scenario described above, the statistics
of multiple successive measurements made on a dynamical system cannot necessarily
be derived from any single density matrix.  Roughly speaking, this is because
every measurement disturbs the state of the system~\cite{Fuch96} in a manner
that depends on the measurement outcome, and because the evolution of the
system state between measurements depends on the full details of the system's
equations of motion.

The effects of repeated measurements on otherwise-unitary quantum evolution
have been extensively studied in quantum optics, with regard to the dynamics
of open quantum systems~\cite{Gard91,Carm93}.  The configuration most often
treated by such work is that of a small, ``encapsulated'' quantum system
having one or more well-defined input/output channels associated with its
coupling to the physical environment.  This picture naturally suggests a
paradigm in which an experimenter attempts to parametrically
identify the Hamiltonian of the encapsulated quantum system by examining the
response of the output channels to driving stimuli applied to the input
channels~\cite{fn1}.  {\em The task of quantum system identification may then
be equated with that of computing the relative likelihood of an observed
sequence of measurement results $\Xi^*$ as a function of the parameter set
${\vec\theta}$, given the external driving conditions imposed by the
experimenter.}  To the extent that the environmental couplings for the system
are known, quantum trajectory theory~\cite{Carm93,Zoll96} suggests a simple
method for the computation, which I discuss below.  Having a likelihood function
$f({\vec\theta}\,|\,\Xi^*)$, one can use maximum-likelihood or Bayesian
principles~\cite{deGr86,Ljun87} to estimate the parameters ${\vec\theta}$.
Note that it should generally be possible for the experimenter to determine
optimal driving conditions that make the system response maximally sensitive
to the values of ${\vec\theta}$, or indeed to {\em adaptively} change the
driving conditions as the estimation starts to converge~\cite{Wise95}.

To illustrate quantum system identification in a concrete setting, let us
focus on an example with relevance to current experiments in cavity quantum
electrodynamics (QED)---a single two-level atom placed within the mode volume
of a driven, high-finesse optical cavity~\cite{Kimb94,Carm93}.  The strength
of the coherent coupling between atom and cavity mode is parametrized
by the vacuum Rabi frequency $g$, whose value depends on the spatial position
of the atom within the cavity.  For a Fabry-Perot resonator $g(\vec r)=g_0
\cos(2\pi x/\lambda)\exp\left[-(y^2+z^2)/w_0^2\right]$, where $x$ is the
coordinate along the cavity axis and $w_0$ is the gaussian waist of the
TEM$_{00}$ resonator mode.  The specific task I shall consider is that of
estimating $g\in[0,g_0]$, which I suppose to be unknown because the atomic
position is not known.  The measurement procedure will simply be to monitor
the arrival-time statistics of photons emitted by the atom-cavity system for
a fixed cavity driving field.  For the purposes of this discussion I
shall not explicitly treat the atom's external degrees of freedom, imagining
that they are fixed by an rf Paul trap or similar confining
mechanism~\cite{Monr95}.  However, note that the correlation of $g$ with the
atomic position {\em operator} implies that ``online'' estimation of $g$ for an
untrapped atom drifting through a cavity could be viewed as a time-distributed
quantum measurement of the position of a free mass~\cite{Cave86,fn1a}.

For a {\it gedanken}-experiment in which the cavity is driven by a resonant
cw probe laser and both the atomic fluoresence and cavity emission are
continuously monitored by perfect photon-counting detectors~\cite{fn2},
the evolution of the conditional state-vector between photodection events
satisfies the effective Schr\"{o}dinger equation $(\hbar=1)$
\begin{eqnarray}
\vert\psi_c(t+dt)\rangle&=&e^{-i{\cal H}_{\rm eff}dt}\vert\psi_c(t)
\rangle,\\
{\cal H}_{\rm eff}(g)&=&ig\left(a\sigma_+ - a^\dagger\sigma_-\right) +
i\epsilon\left(a - a^\dagger\right) - i\kappa a^\dagger a -
i\gamma_\perp\sigma_+\sigma_ .\label{eq:Heff}
\end{eqnarray}
This interaction-picture expression for ${\cal H}_{\rm eff}(g)$ is valid
under the rotating-wave and electric-dipole approximations, and for
identical atomic/cavity/probe-laser frequencies.  Here $\kappa$ is the
field decay rate of the cavity, $\gamma_\perp$ is the dipole decay rate
of the atom, and $\epsilon$ represents the strength of the coherent driving
field.  The jump operator associated with the detection of photons
spontaneously emitted by the atom is ${\hat c}_0\equiv\sqrt{2\gamma_\perp}
\sigma_-$, and ${\hat c}_1\equiv\sqrt{2\kappa}a$ is the jump operator
associated with the detection of photons leaking through the cavity
mirrors.

By registering the origins $\{j_1,\ldots,j_n\}$ ($=0$ for spontaneous
emission or 1 for cavity decays) and arrival times $\{t_1,\ldots,t_n\}$ of
every photon emitted by the atom-cavity system in response to the cw
driving field during an observation interval $[t_0,t_f]$, the hypothetical
experimenter accumulates a classical record $\Xi^*\equiv
(t_0,t_f,\{j_i,t_i\})$ of the stochastic evolution of the system state.
Assuming a uniform prior distribution on $g$, the likelihood function
$f(g\,\vert\,\Xi^*)$ then simply corresponds to a normalized version of
the exclusive probability density~\cite{Carm93,Zoll96}
\begin{equation}
p\left(\Xi^*|\,g\right)={\rm Tr\,}\left[
\begin{array}{c}
U_{\rm eff}\left(t_f,t_n\,;g\right){\hat c}_{j_n}\,
U_{\rm eff}\left(t_n,t_{n-1}\,;g\right)\,\cdots\\
\times\,{\hat c}_{j_1}U_{\rm eff}\left(t_1,t_0\,;g\right)\,\rho(t_0)\,
\;U_{\rm eff}^\dagger\left(t_1,t_0\,;g\right)\,{\hat c}_{j_1}^\dagger\\
\times\,\cdots\,U_{\rm eff}^\dagger\left(t_n,t_{n-1};g\right)\,
{\hat c}_{j_n}^\dagger\,U_{\rm eff}^\dagger\left(t_f,t_n;g\right)
\end{array}\right]\,,\label{eq:EPD}
\end{equation}
{\em viewed as a function of $g$ rather than $\Xi^*$}.  Accordingly,
the maximum-likelihood estimate (MLE) of $g$ is obtained by computing the
value of $g$ which maximizes~(\ref{eq:EPD}) with $\Xi^*$ fixed by the
observed data.  Here $U_{\rm eff}\left(t^\prime,t\,;g\right)$ is the
evolution operator from time $t$ to $t^\prime$ associated with the effective
Hamiltonian ${\cal H}_{\rm eff}(g)$ defined in equation~(\ref{eq:Heff}).

In order to numerically demonstrate quantum system identification
using~(\ref{eq:EPD}), I have generated a set of classical records by
quantum Monte Carlo simluation~\cite{TanPC} of a driven atom-cavity
system with $(g_0,\gamma_\perp,\kappa)/2\pi=(57,2.5,30)$ MHz, and three
different powers for the driving field $\epsilon=\{24,34,44.3\}$. While
this value for $g_0$ is larger than what has been achieved
experimentally~\cite{Thom92}, it should certainly be within reach of works
in progress.  For the simulations I chose an arbitrary atomic position such
that $g({\vec r})=45$ MHz, and generated classical records with an
observation time of 1 $\mu$s each.  Figure~1a illustrates the stochastic
time-evolution of the mean intracavity photon number, taken from typical
Monte-Carlo data sets for each of the three values of $\epsilon$.  The
photocount statistics are clearly super-Poissonian, and the simulated data
show that quantum jumps often occur at a local rate that greatly
exceeds the rate at which the system regresses to steady state.

For each Monte Carlo trajectory, an identification routine based
on~(\ref{eq:EPD}) was used to compute the stochastic time-evolution of
$f(g\,\vert\,\Xi^*)$, as well as the corresponding MLE.  Figure~1b
shows one typical data set with $\epsilon=34$, starting from the inital
estimate made after only one photodetection event and updated after each
subsequent photodetection.  Figure~2a indicates the ensemble-averaged
convergence of the MLE for $g$, based on 2000 simulations for $\epsilon=24$,
300 simulations for $\epsilon=34$, and 150 simulations for $\epsilon=44.3$.
A histogram representing the time-evolution of the MLE sampling
distribution is given in Figure~2b, for the case of $\epsilon=44.3$.
With this driving field, $\sim 1\%$ accuracy in estimation of $g$ is obtained
in 1 $\mu$s observation time ($\sim 600$ quantum jumps).

It is important to note that the QPE procedure described above automatically
makes efficient use of any information about $g$ that is contained in
higher-order correlations of the classical record of counting times.  Of
course, not every open quantum system will generate significant correlations
of this type.  In the scenario discussed above for example, the photon stream
emitted by the atom-cavity system would become nearly Poissonian in the limit
of either weak excitation (corrleated pairs of photons become rare) or of weak
coupling $g\ll\kappa,\gamma_\perp$ (correlations become weak). Two critical
conditions for correlations to be strongly evident in individual classical
records are that the mean time between counts must be comparable to or less
than the system regression time~\cite{Rice88}, and that the system dynamics
must be significantly altered by the loss of a single quantum of
excitation~\cite{Remp91}.  For systems not satisfying these criteria, the
methods described above offer no real advantage over statistical estimators
based on only the {\em steady-state} density matrix obtained by solving the
master equation associated with~(\ref{eq:Heff}).  Accordingly, optimal parameter
estimation in such systems can be formulated within the paradigm of conventional
QPE.  Judging from the trend shown in Figure~2a however, it certainly seems that
the information rate on parameters in strongly-coupled systems can be
significantly larger in the strong-driving regime~\cite{Alsi91,fn3} than in
the weak-field regime.

In closing, let me note that a straightforward extension of the above method
would allow the identification of non-stationary Hamiltonians in which the
parameters ${\vec\theta}(t)$ vary slowly compared to the timescale for
convergence of the corresponding statistical estimators.  Relative to the
example discussed above, a recent cavity-QED experiment incorporating a
laser-cooled atomic source~\cite{Mabu96} has demonstrated the practical
feasibility of achieving this separation of timescales.  It seems reasonable
to hope that the methods proposed above could be utilized in future
experimental work to track variations in $g$ associated with the motion of
an individual atom through the mode-volume of a high-finesse optical
cavity~\cite{fn2}.  A digital signal processor implementing such a procedure
could be used as the state observer in a ``semiclassical'' feedback control
loop designed to confine and cool the atom's center-of-mass motion.

I wish to acknowledge invaluable conversations with H. J. Kimble, G. J.
Milburn, S. M. Tan, Q. A. Turchette, D. W. Vernooy, and P. Zoller, as well
as the support of a National Defense Science and Engineering Graduate
Fellowship.

%\Figures

\begin{figure}
\caption{(a) Time-evolution of the mean intracavity photon number $\langle
a^\dagger a\rangle$ in
individual trajectories.  Top trace (i) is for $\epsilon=24$, middle
trace (ii) is for $\epsilon=34$, and bottom trace (iii) is for
$\epsilon=44.3$.
(b) Corresponding stochastic evolution of the (normalized) likelihood
function $f(g\vert\Xi^*)$ and corresponding MLE in one quantum trajectory
with driving field amplitude $\epsilon=34$.  The surface height indicates
relative probability of $g\in[35,57]$, with a resolution of 1.  The ``true''
value of $g$ corresponds to 45.  Note that the likelihood function is updated
each time a photon is detected, so that the timelike coordinate in this
surface plot corresponds to jump number rather than absolute time.}
\end{figure}

\begin{figure}
\caption{(a) Standard deviation of the maximum-likelihood estimator for
$g$ as a function of absolute time ($+$---$\epsilon=24$,
$\circ$---$\epsilon=34$, $\times$---$\epsilon=44.3$). (b) Histogram
showing the evolution (in absolute time) of the sampling distribution for
the MLE of $g$, representing 150 simulations with $\epsilon=44.3$.}
\end{figure}

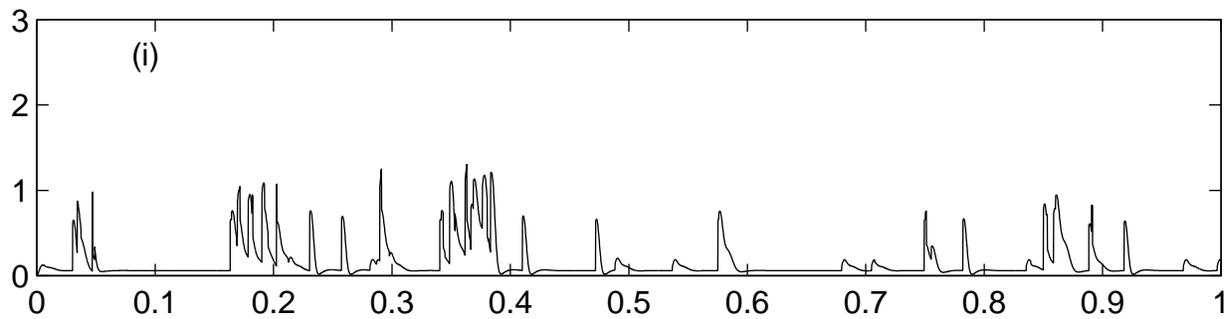
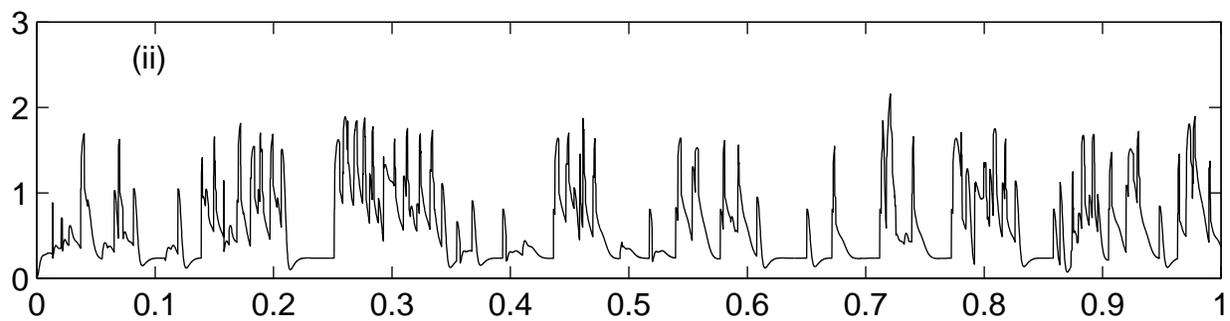
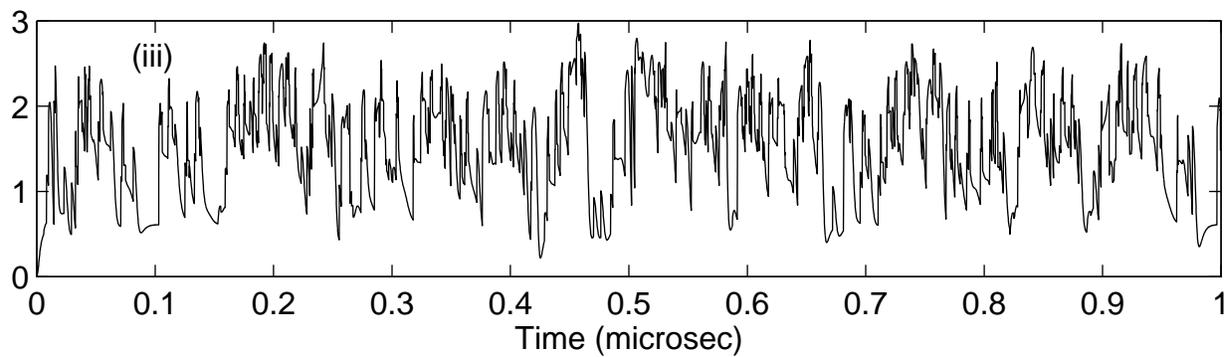
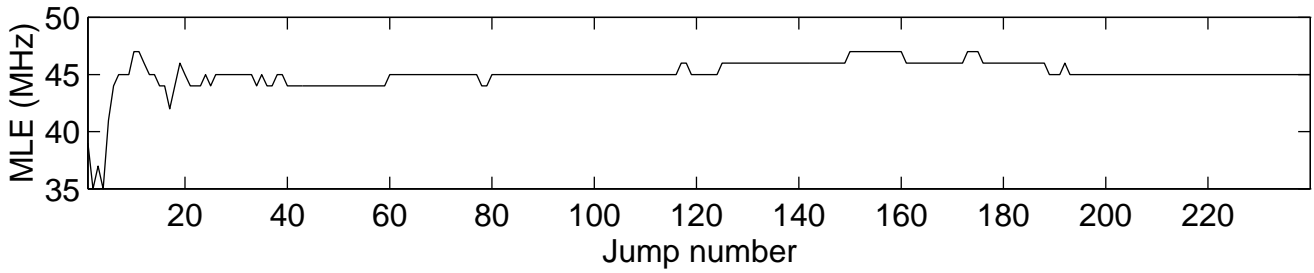
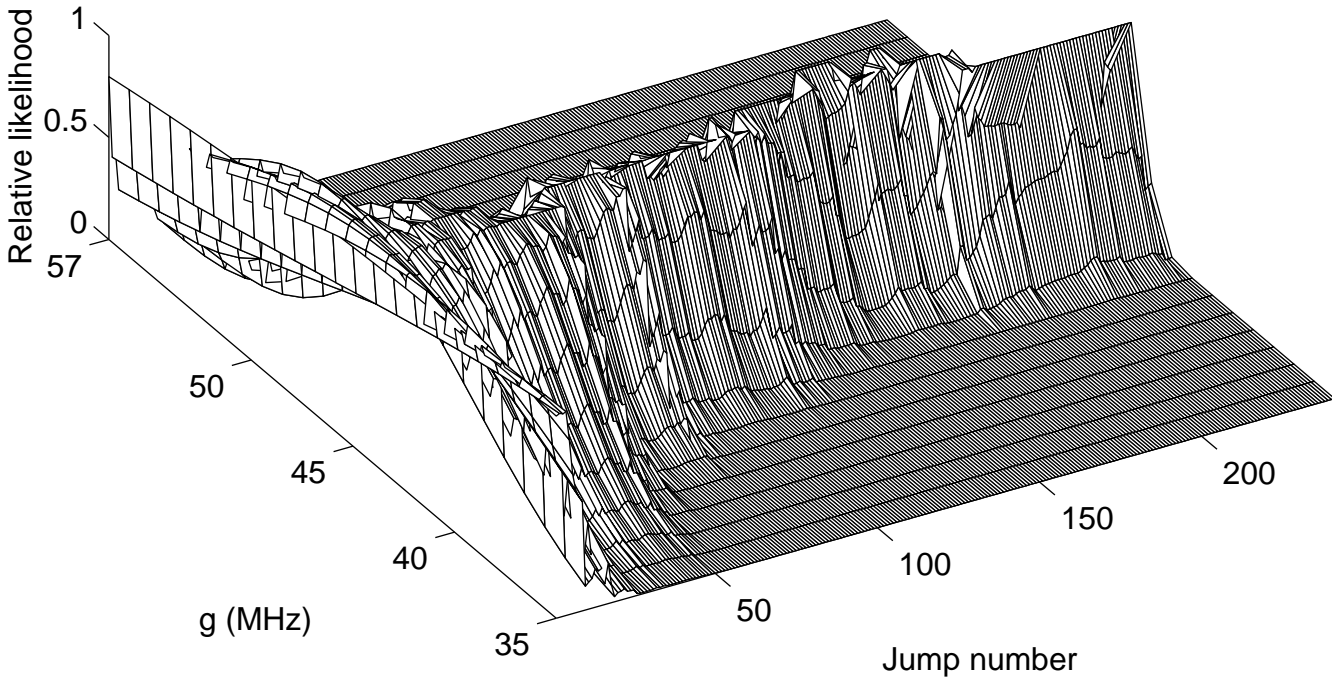
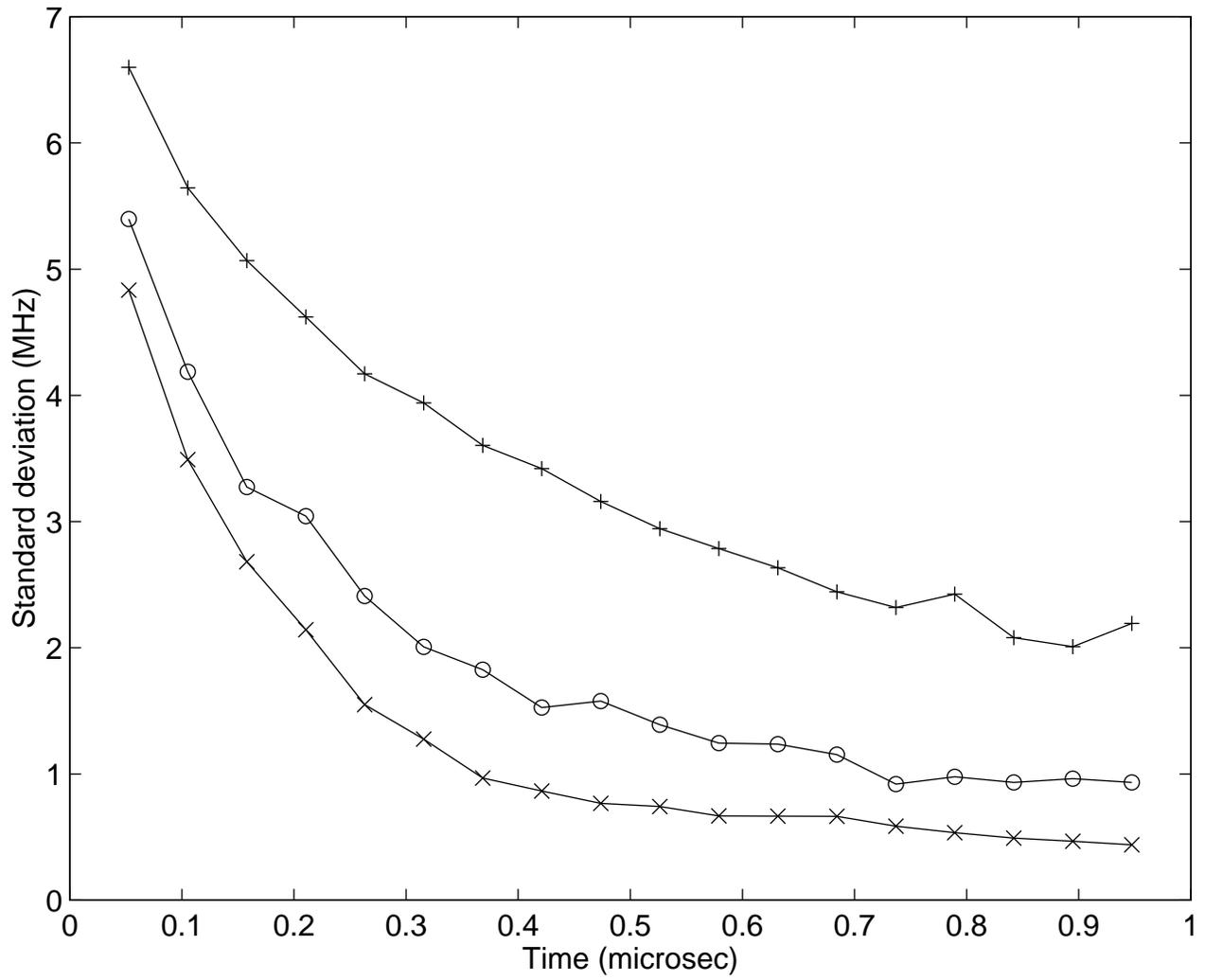
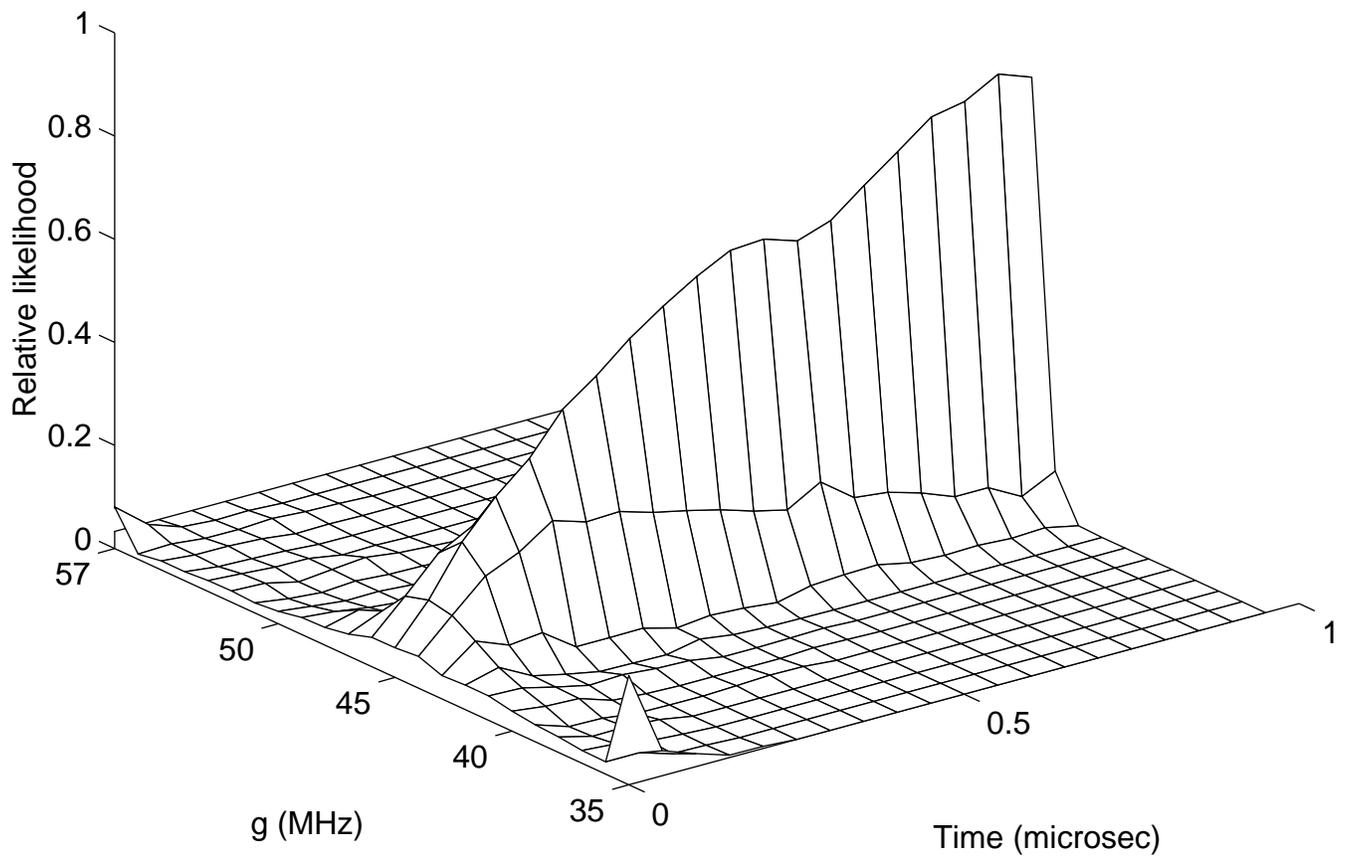
\end{document}